\documentclass{article}
\usepackage{wrapfig}
\usepackage{natbib}
\bibpunct{(}{)}{,}{a}{,}{;}

\usepackage{amsmath,amsfonts,amsthm,amssymb,latexsym}
\usepackage[pdftex]{graphicx}
\usepackage{natbib}

\newcommand\SI{{\em Statistical Inference}}

\newenvironment{kvothe}{\small\begin{quote}\begin{em}}{\end{em}\end{quote}\normalsize\vglue -0.05truecm}
\newcommand\dropcap\noindent

\begin{document}

\title{Inherent Difficulties of Non-Bayesian Likelihood-based Inference, as Revealed by an Examination of a Recent Book by Aitkin\footnote{
  {\sf gelman@stat.columbia.edu}, {\sf xian@ceremade.dauphine.fr}, {\sf rousseau@ensae.fr}
}}
\author{{\sc A. Gelman${}^1$, C.P.~Robert${}^{2,3,4}$, and J. Rousseau${}^{2,4,5}$}\\
  ${}^1$Depts. of Statistics and of Political Science, Columbia University\\ 
  ${}^2$Universit\'e Paris-Dauphine, CEREMADE\\
  ${}^3$Institut Universitaire de France, ${}^4$CREST, and ${}^5$ENSAE\\ 
}       

\maketitle

\begin{abstract}
For many decades, statisticians have made attempts to prepare the Bayesian omelette without breaking the Bayesian eggs; that is, to obtain probabilistic likelihood-based inferences without relying on informative prior distributions.  A recent example is Murray Aitkin's recent book, {\em Statistical Inference}, which presents an approach to statistical hypothesis testing based on comparisons of posterior distributions of likelihoods under competing models. Aitkin develops and illustrates his method using some simple examples of inference from iid data and two-way tests of independence.  We analyze in this note some consequences of the inferential paradigm adopted therein, discussing why the approach is incompatible with a Bayesian perspective and why we do not find it relevant for applied work. 
\end{abstract}


\section{Introduction}

For many decades, statisticians have made attempts to prepare the Bayesian omelette without breaking the
Bayesian eggs; that is, to obtain probabilistic likelihood-based inferences without relying on informative
prior distributions.  A recent example is Murray Aitkin's recent book, {\em Statistical Inference}, which is
the culmination of a long research program on the topic of integrated evidence, exemplified by the discussion
paper of \cite{aitkin:1991}.  The book, subtitled {\em An Integrated Bayesian/Likelihood Approach}, proposes
handling statistical hypothesis testing and model selection via comparisons of posterior distributions of
likelihood functions under the competing models or via the posterior distribution of the likelihood ratios
corresponding to those models. (The essence of the proposal is detailed in Section \ref{small?}.) Instead of
comparing Bayes factors or performing posterior predictive checks (comparing observed data to posterior
replicated pseudo-datasets), \SI~recommends a fusion between likelihood and Bayesian paradigms that allows for
the perpetuation of noninformative priors in testing settings where standard Bayesian practice prohibits their
usage \citep{degroot:1973} or requires an extended decision-theoretic framework \citep{bernardo:2011}. While we
appreciate the considerable effort made by Aitkin to place his theory within a Bayesian framework, we remain
unconvinced of the said coherence, for reasons exposed in this note.

\begin{wrapfigure}{l}{0.22\textwidth}
  \begin{center}
  \includegraphics[width=0.18\textwidth]{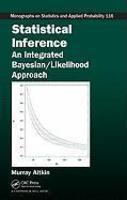}
  \end{center}
  \caption{{\it Cover of \SI}}
\end{wrapfigure}

From our Bayesian perspective, and for several distinct reasons detailed in the present note, integrated
Bayesian/likelihood inference cannot fit within the philosophy of Bayesian inference. Aitkin's commendable
attempt at creating a framework that incorporate the use of arbitrary noninformative priors in model choice
procedures is thus incoherent in this Bayesian respect. When using improper priors lead to meaningless Bayesian
procedures for posterior model comparison, we see this as a sign that the Bayesian model will not work for the
problem at hand.  Rather than trying at all cost to keep the offending model and define marginal posterior
probabilities by fiat (whether by BIC, DIC, intrinsic Bayes factors, or posterior likelihoods), we prefer to
follow the full logic of Bayesian inference and recognize that, when one's Bayesian approach leads to a
dead end, one must change either one's methodologies or one's beliefs (or both).  Bayesians, both subjective and
objective, have long recognized the need for tuning, expanding, or otherwise altering a model in light of its
predictions (see, for example, \citealp{good:1950} and \citealp{jaynes:2003}), and we view undefined Bayes
factors as an example where otherwise useful methods are being extended beyond their applicability.
To try to work around such problems without altering the prior distribution is, we believe, an abandonment of
Bayesian principles and, more importantly, an abandoned opportunity for model improvement.

The criticisms found in the current review are therefore not limited to Aitkin's book; they also
apply to previous patches such as the deviance information criterion (DIC) of \cite{spiegbestcarl} (which also
uses a ``posterior" expectation of the log-likelihood) and the pseudo-posteriors of
\cite{geisser:eddy:1979}(which make an extensive use of the data in their product of predictives). 

Unlike the author, who has felt the call to construct a partly new \citealp{aitkin:1991} if tentatively
unifying foundation for statistical inference, we have the luxury of feeling that we already live in a
comfortable (even if not flawless) inferential house.  Thus, we come to Aitkin's book not with a perceived need
to rebuild but rather with a view toward strengthening the potential shakiness of the pillars that support our
own inferences. A key question when looking at any method for probabilistic inference that is not fully
Bayesian is:  For the applied problems that interest us, does the proposed new approach achieve better
performances than our existing methods?  Our answer, to which we arrive after careful thought, is no.


As an evaluation of the ideas found in \SI, the criticisms found in this review are inherently limited.  We do
not claim here that Aitkin's approach is wrong {\em per se} 
merely that
it does not fit within our inferential methodology, namely Bayesian statistics, despite using Bayesian tools.
We acknowledge that statistical methods do not, and most likely never will, form a seamless logical structure.
It may thus very well be that the approach of comparing posterior distributions of likelihoods could be useful
for some actual applications, and perhaps Aitkin's book will inspire future researchers to demonstrate this.

\SI~begins with a crisp review of frequentist, likelihood and Bayesian approaches to inference and then
proceeds to the main issue: introducing the ``integrated Bayes/like\-li\-hood approach'', first described in
Chapter 2.  Much of the remaining methodological material appears in Chapters 4 (``Unified analysis of finite
populations'') and 7 (``Goodness of fit and model diagnostics'').  The remaining chapters apply Aitkin's
principles to various examples.  The present article discusses the basic ideas in \SI, then consider the
relevance of Aitkin's methodology within the Bayesian paradigm.

\section{A small change in the paradigm}\label{small?}
\subsection{Posterior likelihood}\label{post/like}
\begin{kvothe} ``This quite small change to standard Bayesian analysis allows a very general approach to a wide range of
apparently different inference problems; a particular advantage of the approach is that it can use the same
noninformative priors." \SI, page xiii\end{kvothe}

The ``quite small change" advocated by  \SI~consists in envisioning the likelihood function $L(\theta,x)$ 
as a generic function of the parameter $\theta$ that can be processed a posteriori (that is, with a distribution induced by
the posterior $\pi(\theta|x)$), hence allowing for (posterior) cdf, mean, variance and quantiles. In particular, the
central tool for Aitkin's model fit is the ``posterior cdf" of the likelihood,
$$
F(z) = \mbox{Pr}^\pi(L(\theta,x)>z|x)\,.
$$ 
As argued by the author (Chapter 2, page 21), this ``small change" in perspective has several appealing features:
\begin{itemize}
\item[--] The approach is general and allows to resolve the difficulties with the Bayesian processing of point
null hypotheses, being defined solely by the Bayesian model associated with $L(\theta,x)$;
\item[--] The approach allows for the use of generic noninformative and improper priors, again by being relative
to a single model;
\item[--] The approach handles more naturally the ``vexed question of model fit", 
still for the same
reason;
\item[--] The approach is ``simple.'' 
\end{itemize}
As noted above, the setting is quite similar to Spiegelhalter et al.'s (\citeyear{spiegbestcarl}) DIC in that 
the deviance $D(\theta)=-2 \log(f(x|\theta))$ is a renaming of the likelihood and is considered ``a posteriori"
both in $\bar{D}=\mathbb{E}[D(\theta)|x]$ and in $p_D=\bar{D}-D(\hat{\theta})$, where $\hat{\theta}$
is a Bayesian estimator of $\theta$, since
$$
    \mathit{DIC} = p_D+\bar{D}. 
$$
The discussion of \cite{spiegbestcarl} made this point clear, see in particular \cite{dawid:2002}, even though
the authors disagreed. \cite{plummer:2008} make a similarly ambiguous proposal that also relates to
\cite{geisser:eddy:1979} by its usage of cross-validation quantities.

We however dispute both the appropriateness and the magnitude of the change advocated in \SI~and show below
why, in our opinion, this shift in paradigm constitutes a new branch of statistical inference, differing from
Bayesian analysis on many points. First, using priors and posteriors is no guarantee that inference is Bayesian
\citep{Seidenfeld:1992}. Empirical Bayes techniques are witnesses of this
\citep{robbins:1964,carlin:louis:2008}.  Aitkin's key departure from Bayesian principles means that his
procedure has to be validated on its own, rather than benefiting from the coherence inherent to Bayesian
procedures. The practical advantage of the likelihood/Bayesian approach may be convenience,
but the drawback is that the method pushes both the user and the statistician {\em away} from progress in model
building.\footnote{One might argue that, in practice, almost all Bayesians are subject to our criticism of
``using models that make nonsensical predictions.''  For example, \cite{gelman:carlin:stern:rubin:2001} and
\cite{marin:robert:2007} are full of noninformative priors.  Our criticism here, though, is not of
noninformative priors in general but of incoherent predictions about {\em quantities of interest}.  In
particular, noninformative priors can often (but not always!) give reasonable inferences about parameters
$\theta$ within a model, even while giving meaningless (or at least not universally accepted) values for
marginal likelihoods that are needed for Bayesian model comparison.  It does when interest shifts from
$\mbox{Pr}(\theta|x,H)$ to $\mbox{Pr}(H|x)$ that the Bayesian must set aside most of noninformative
$\pi(\theta|H)$ and, perhaps reluctantly, set up an informative model. See, e.g., \cite{liang:etal:2008} and
\cite{johnson:rossell:2010} for some current perspectives on Bayesian model choice using noninformative
priors.}

We envision Bayesian data analysis as comprising three steps:  (1) model building, (2) inference, and (3) model
checking.  In particular, we view steps (2) and (3) as separate.  Inference works well, with many exciting
developments still in the coming, handling complex models, leading to an unlimited range of applications, and a
partial integration with classical approaches (as in the empirical Bayes work of \citealp{efron:morris:1975},
or more recently the similarities between hierarchical Bayes and frequentist false discovery rates discussed by
\citealp{efron:2010}), causal inference, machine learning, and other aims and methods of statistical inference.

Even in the face of all this progress on inference, Bayesian model checking remains a bit of an anomaly, with
the three leading Bayesian approaches being Bayes factors, posterior predictive checks, and comparisons of
models based on prediction error and other loss-based measures. (Decision-theoretic analyses as in
\citep{bernardo:2011}, while intellectually convincing, have not gained the same amount of popularity.)
Unfortunately, as Aitkin points out, none of these model checking methods works completely smoothly:  Bayes
factors depend on aspects of a model that are untestable and are commonly assigned arbitrarily; posterior
predictive checks are, in general, ``conservative" in the sense of producing $p$-values whose probability
distributions are concentrated near $0.5$; and prediction error measures (which include cross-validation and
DIC) require the user to divide data into test
and validation sets, lest they use the data twice (a point discussed immediately below). The setting is even bleaker when
trying to incorporate noninformative priors \citep{gelman:carlin:stern:rubin:2001,robert:2001} and new
proposals are clearly of interest.

\subsection{``Using the data twice"}
\begin{kvothe}
``A persistent criticism of the posterior likelihood approach (...) has been based on the claim that these approaches are
`using the data twice,' or are `violating temporal coherence." \SI, page 48\end{kvothe}

``Using the data twice" is not our main reservation about the method---if only because this is a
rather vague concept. Obviously, one could criticize the use of the ``posterior expectation" of the likelihod
as being the ratio of the marginal of the twice replicated data over the marginal of the original data, 
$$
\mathbb{E}[L(\theta,x)|x] = \int L(\theta,x) \pi(\theta|x)\,\text{d}\theta = \dfrac{m(x,x)}{m(x)}\,,
$$
similar to \cite{aitkin:1991} (a criticism clearly expressed in the discussion therein). However, a more
fundamental issue is that the ``posterior" distribution of the likelihood function cannot be justified from a
Bayesian perspective.  \SI~stays away from decision-theory (as stated on page xiv) so there is no derivation
based on a loss function or such. Our primary difficulty with the integrated likelihood idea (and DIC as well)
is (a) that the likelihood function does not exist a priori and (b) that it requires a joint distribution to be
properly defined in the case of model comparison.  The case for (a) is arguable, as Aitkin would presumably
contest that there exists a joint distribution on the likelihood, even though the case of an improper prior
stands out (see below). We still see the concept of a posterior probability that the likelihood ratio is larger
than $1$ as meaningless. The case for (b) is more clear-cut in that when considering two models, hence a
likelihood ratio, a Bayesian analysis does require a joint distribution on the two sets of parameters to reach
a decision, even though in the end only one set will be used.  As detailed below in Section \ref{ProdPost},
this point is related with the introduction of pseudo-priors by \cite{carlin:chib:1995} who needed arbitrary
defined prior distributions on the parameters that do not exist.

In the specific case of an improper prior, Aitkin's approach cannot be validated in a probability setting for
the reason that there is no joint probability on $(\theta,x)$.  Obviously, one could always advance that the
whole issue is irrelevant since improper priors do not stand within probability theory.  However, improper
priors do stand within the Bayesian framework, as demonstrated for instance by \cite{hartigan:1983} and it is
easy to give those priors an exact meaning. When the data are made of $n$ iid observations
$x^n=(x_1,\ldots,x_n)$ from $f_\theta$ and an improper prior $\pi$ is used on $\theta$, we can consider a {\em
training sample} \citep{smith:spiegelhalter:1982} $x^{(l)}$, with $(l) \subset \{ 1,...,n\}$  such that 
$$ \int f(x^{(l)}|\theta)\,\text{d}\pi(\theta) < \infty \qquad (l\leq n). $$
If we construct a probability distribution on $\theta$ by 
$$\pi_{x^{(l)}}(\theta) \propto \pi(\theta)f(x^{(l)}|\theta)\,,$$
the posterior distribution associated with this distribution and the remainder of the sample $x^{(-l)}$ is
given by 
$$\pi_{x^{(l)}}(\theta|x^{(-l)}) \propto \pi(\theta)f(x^n|\theta), \quad x^{(-l)} = \{ x_i , i \notin (l)\}\,.
$$
This distribution is independent from the choice of the training sample; it only depends on the likelihood of
the whole data $x^n$ and it therefore leads to a non-ambiguous posterior distribution\footnote{ Obvious
extensions to the case of independent but non iid data or of exchangeable data lead to the same interpretation.
The case of dependent data is more delicate, but similar interpretation can still be considered.} on $\theta$.
However, as is well known, this construction does not lead to produce a joint distribution on $(x^n, \theta)$,
which would be required to give a meaning to Aitkin's integrated likelihood. Therefore, his approach cannot
cover the case of improper priors within a probabilistic framework and thus fails to solve the very difficulty
with noninformative priors it aimed at solving. This is further illustrated by the use of Haldane's prior
in Chapter 4 of \SI, despite it not allowing for empty cells in a contingency table \citep{jeffreys:1939}.

\section{Posterior probability on the posterior probabilities}
\begin{kvothe} ``The $p$-value is equal to the posterior probability that the
likelihood ratio, for null hypothesis to alternative, is greater than 1 (...)
%
The posterior probability is $p$ that the posterior probability of $H_0$ is greater than 0.5.'' \SI, pages 42--43
\end{kvothe}

Those two equivalent statements show that it is difficult to give a Bayesian interpretation to Aitkin's method,
since the two ``posterior probabilities" quoted above are incompatible. Indeed, a fundamental Bayesian property
is that the posterior probability of an event related with the parameters of the model is not a random quantity
but a number. To consider the ``posterior probability of the posterior probability" means we are exiting the
Bayesian domain, both from logical and philosophical viewpoints.

In Chapter 2, Aitkin exposes his (foundational) reasons for choosing this new approach by integrated
Bayes/likelihood. His criticism of Bayes factors is based on several points we feel useful to reproduce here:
\begin{enumerate}
\renewcommand{\theenumi}{(\roman{enumi})}

\item\label{i} ``Have we really eliminated the uncertainty about the model parameters by integration?  The
integrated likelihood (...) is the expected value of the likelihood. But what of the prior variance of the
likelihood?" (page 47).

\item\label{ii} ``Any expectation with respect to the prior implies that the data has not yet been observed
(...) So the ``integrated likelihood" is the joint distribution of random variables drawn by a two-stage
process. (...) The marginal distribution of these random variables is not the same as the distribution of $Y$
(...) and does not bear on the question of the value of $\theta$ in that population" (page 47).

\item\label{iii} ``We cannot use an improper prior to compute the integrated likelihood. This eliminate the
usual improper noninformative priors widely used in posterior inference." (page 47).

\item\label{iv} ``Any parameters in the priors (...) will affect the value of the integrated likelihood and
this effect does not disappear with increasing sample size" (page 47).

\item\label{v} ``The Bayes factor is equal to the posterior mean of the likelihood ratio between the models"
{\em [meaning under the full model posterior]} (page 48).

\item\label{vi} ``The Bayes factor diverges as the prior becomes diffuse. (...) This property of the Bayes
factor has been known since the Lindley/Bartlett paradox of 1957'' (page 48).  \end{enumerate}

The representation \ref{i} of the ``integrated" (or marginal) likelihood as an expectation under the prior 
$$
m(x) = \int L(\theta,x) \pi(\theta)\,\text{d}\theta = \mathbb{E}^{\pi}[ L(\theta,x) ]
$$
is unassailable and is for instance used as a starting point for motivating the nested sampling method
\citep{skilling:2007a,chopin:robert:2010}. This does not imply that the extension to the variance or to any
other moment stated in \ref{i} has a similar meaning, nor that the move to the expectation under the posterior is valid within
the Bayesian paradigm.  While the difficulty \ref{iii} with improper priors is real, and while the impact of
the prior modelling \ref{iv} may have a lingering effect, the other points can be easily rejected on the
ground that the posterior distribution of the likelihood is meaningless within a Bayesian perspective.  This
criticism is anticipated by Aitkin who protests on pages 48-49 that, given point \ref{v}, the posterior
distribution must be ``meaningful," since the posterior mean is ``meaningful", but the interpretation of the
Bayes factor as a ``posterior mean" is only an interpretation of an existing integral (in the specific case of
nested models), it does not give any validation to the analysis.  (The marginal likelihood may similarly be
interpreted as a prior mean, despite depending on the observation $x$, as in the nested sampling perspective.
More generaly, bridge sampling techniques also exploit those multiple representations of a ratio of integrals,
\citealp{gelman:meng:1998}.)
One could just as well take \ref{ii} above as an argument {\em against} the integrated likelihood/Bayes
perspective.

\section{Products of posteriors}\label{ProdPost}

\begin{figure}[hbtp]
\centering
\includegraphics[width=.95\textwidth]{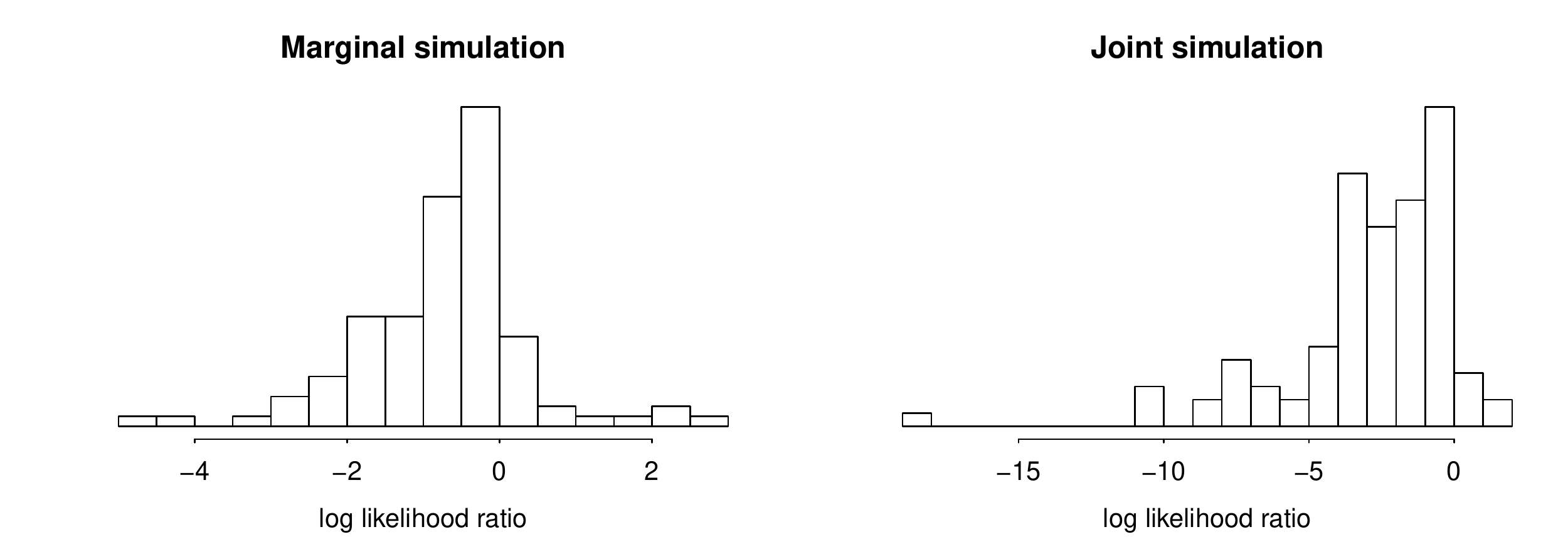}
\caption{\label{fig:sXot}
{\it Comparison of the distribution of the likelihood ratio under the correct joint posterior
and under the product of the model-based posteriors, when assessing a Poisson model against a
binomial with $m=5$ trials, for $x=3$. The joint simulation produces
a much more supportive argument in favor of the negative binomial model, when
compared with the product of the posteriors.} 
}
\end{figure}

In the case of unrelated models to be compared, the fundamental theoretical argument against using posterior
distributions of the likelihoods and of related terms is that the approach leads to parallel and separate simulations from
the posteriors under each model. 
\SI~recommends that models be compared via the distribution of the likelihood ratio values,
$${L_i(\theta_i|x)}\bigg/{L_k(\theta_k|x)},$$
where the $\theta_i$'s and $\theta_k$'s are drawn from the respective posteriors. This choice is similar to
Scott's (\citeyear{scott:2002}) and to Congdon's (\citeyear{congdon:2006}) mistaken solutions exposed in
\cite{robert:marin:2008}, in that MCMC simulations are run for each model separately and the resulting samples
are then gathered together to produce either the posterior expectation (in Scott's, 2002, case) or the posterior
distribution (for the current paper) of
$$\rho_i L(\theta_i|x) \bigg/ \sum_k \rho_k L(\theta_k|x)\,,$$
which do not correspond to genuine Bayesian solutions (see \citealp{robert:marin:2008}). Again, this is not as much
because the dataset $x$ is used repeatedly in this process (since reversible MCMC produces as well separate samples from the
different posteriors) as the fundamental lack of a common joint distribution that is needed in the Bayesian framework. This
means, e.g., that the integrated likelihood/Bayes technology is producing samples from the product of the posteriors (a
product that clearly is not defined in a Bayesian framework) instead of using pseudo-priors as in \cite{carlin:chib:1995}, 
i.e. of considering a joint posterior on $(\theta_1,\theta_2)$, which is [proportional to]
\begin{equation}\label{eq:joint}
p_1 m_1(x) \pi_1(\theta_1|x) \pi_2(\theta_2)+p_2 m_2(x) \pi_2(\theta_2|x) \pi_1(\theta_1).
\end{equation}
This makes a difference in the outcome, as illustrated in Figure \ref{fig:sXot}, which
compares the distribution of the likelihood ratio under the true posterior and under the product of posteriors, when  
assessing the fit of a Poisson model against the fit of a binomial model with $m=5$ trials, for the observation $x=3$. 
The joint simulation produces a much more supportive argument in favor of the binomial model, when compared 
with the product of the posteriors. (Again, this is inherently the flaw found in the reasoning leading to
Scott's, 2002, and Congdon's, 2006, methods for approximating Bayes factors.)

Although we do not advocate its use, a Bayesian version of Aitkin's proposal can be constructed based on the
following loss function that evaluates the estimation of the model index $j$ based on the
values of the parameters under both models and on the observation $x$:
\begin{eqnarray} \label{loss}
L(\delta, (j,\theta_j,\theta_{-j})) = \mathbb{I}_{\delta=1} \mathbb{I}_{f_2(x | 
\theta_2)> f_1(x|\theta_1)} + \mathbb{I}_{\delta= 2} 
\mathbf{I}_{f_2(x | \theta_2)< f_1( x |\theta_1)} \,.
\end{eqnarray}
Here $\delta = j$ means that model $j$ is chosen, and $f_j(.|\theta_j) $ denotes the likelihood under model $j$. 
Under this loss, the Bayes (optimal) solution is 
$$
\delta^\pi(x) = \begin{cases} 1 &\mbox{if } \mbox{Pr}^\pi\left[ f_2( x  | \theta_2)< f_1( x |\theta_1)| x  \right] > 1/2\\
2 &\text{otherwise,}
\end{cases}
$$
which  depends on the {\em joint} posterior distribution \eqref{eq:joint} on $(\theta_1,\theta_2)$, thus differs from
Aitkin's solution. We have
\begin{align*}
\mbox{Pr}^\pi\left[ f_2( x  | \theta_2)< f_1( x |\theta_1)| x  \right]  
=& \pi(\mathcal M_1| x  ) \int_{\Theta_2}  \mbox{Pr}^{\pi_1}\left[ l^1(\theta_1) > l^2(\theta_2)| 
x, \theta_2 \right] \,\text{d}\pi_2(\theta_2)\\
&+ \pi(\mathcal M_2| x  ) \int_{\Theta_1} \mbox{Pr}^{\pi_2}\left[ l^1(\theta_1) > l^2(\theta_2)| 
x , \theta_1\right] \,\text{d}\pi_1(\theta_1)\,,
\end{align*}
where $l^1$ and $l^2$ denote the log-likelihoods and where the probabilities within the integrals 
are computed under $\pi_1(\theta_1|x)$ and $\pi_2(\theta_2|x)$, respectively.
(Pseudo-priors as in \citealp{carlin:chib:1995} could be used instead of the true priors, a requirement when 
at least one of those priors is improper.)

An asymptotic evaluation of the above procedure is possible: consider a sample of size $n$, $x^n$. 
If $\mathcal M_1$ is the ``true" model, then
$\pi(\mathcal M_1|x^n ) = 1+ o_p(1)$ and we have 
\begin{align*}
\mbox{Pr}^{\pi_1}\left[ l^1(\theta_1) > l^2(\theta_2)|x^n , \theta_2\right] &= 
\mbox{Pr} \left[ -\mathcal X^2_{p_1} > l^2(\theta_2)- l^2(\hat{\theta_1}) \right]  + O_p(1/\sqrt{n}) \\
&= F_{p_1}\left[ l^1(\hat{\theta_1}) - l^2(\theta_2) \right] + O_p(1/\sqrt{n})\,,
\end{align*}
with obvious notations for the corresponding log-likelihoods, $p_1$ the dimension of $\Theta_1$, $\hat\theta_1$ the maximum
likelihood estimator of $\theta_1$, and $\mathcal X^2_{p_1}$ a  chi-square random variable with $p_1$ degrees of freedom. 
Note also that, since $l^2(\theta_2) \leq l^2(\hat{\theta}_2) $,  
$$l^1(\hat{\theta}_1) - l^2(\theta_2) \geq   n\text{KL}(f_0, f_{\theta_2^*}) + O_p(\sqrt{n})\,,$$
where $\text{KL}(f,g)$ denotes the Kullback--Leibler divergence and $\theta^*_2$ denotes the \textit{projection} of the true model on $\mathcal  M_2$ : $\theta^*_2 = \mbox{argmin}_{\theta_2} KL(f_0,f_{\theta_2})$, we have
$$
\mbox{Pr}^\pi\left[ f(x^n | \theta_2)< f(x^n|\theta_1)|x^n \right]  =1 + o_p(1)\,.
$$
By symmetry, the same asymptotic consistency occurs under model $\mathcal M_2$. On 
the opposite, Aitkin's approach leads (at least in regular models) to the approximation
$$
\mbox{Pr}[ \mathcal X^2_{p_2} - \mathcal X^2_{p_1} > l^2(\hat{\theta}_2) - l^1(\hat{\theta}_1)],
$$
where the $\mathcal X^2_{p_2}$ and $ \mathcal X^2_{p_1}$ random variables are independent, hence producing
quite a different result that depends on the asymptotic behavior of the likelihood ratio. Note that for both
approaches to be equivalent one would need a pseudo-prior for $\mathcal M_2$ (resp. $\mathcal M_1$ if $\mathcal
M_2$ were \textit{true}) as tight around the maximum likelihood as the posterior $\pi_2(\theta_2|x^n)$, which
would be equivalent to some kind of empirical Bayes type of procedure. 

Furthermore, in the case of embedded models, $\mathcal M_2$ and $\mathcal M_1
\subset \mathcal M_2$, Aitkin's approach can be given a
probabilistic interpretation. To this effect, we write the parameter under
$\mathcal M_1$ as $(\theta_1, \psi_0)$, $\psi_0$ being a fixed known quantity,
and under $\mathcal M_2 $ as $\theta_2 = (\theta_1, \psi)$, so that comparing
$\mathcal M_1$ with $\mathcal M_2$ corresponds to testing the null hypothesis
$\psi = \psi_0$.  Aitkin does not impose a positive prior probability on
$\mathcal M_1$, since his prior only bears on $\mathcal M_2$ (in a spirit close
to the Savage-Dickey representation, see \citealp{marin:robert:2010b}). His
approach is therefore similar to the inversion of a confidence region into a
testing procedure (or vice-versa).  Under the model $\mathcal M_1 \subset
\mathcal M_2$, denoting by $l(\theta,\psi)$ the log-likelihood of the bigger model,
\begin{eqnarray*}
\mbox{Pr}^\pi\left[ l(\theta_1,\psi_0) > l(\theta_1,\psi) |x^n \right]
&\approx& \mbox{Pr}\left[ \mathcal X^2_{p_2-p_1} > -
l(\hat{\theta}_1(\psi_0), \psi_0)+ l(\hat{\theta}_1,\hat{\psi})\right]  \\
&\approx & 1 - F_{p_2-p_1}[ - l(\hat{\theta}_1(\psi_0), \psi_0)+ l(\hat{\theta}_1,\hat{\psi}) ], 
\end{eqnarray*} 
which is the approximate $p$-value associated with the likelihood ratio test. Therefore, the aim of this
approach seems to be, at least for embedded models where the Bernstein--von Mises theorem holds for the
posterior distribution, to construct a \textit{Bayesian} procedure reproducing the $p$-value associated with
the likelihood ratio test. From a frequentist point of view it is of interest to see that the posterior
probability of the likelihood ratio being greater than one is approximately a $p$-value, at least in cases when
the Bernstein-von Mises theorem holds, e.g.~for embedded models and proper priors. This $p$-value
can then  be given a finite-sample meaning (under the above restrictions), however it seems more interesting
from a frequentist perspective than from a Bayesian one.\footnote{See Chapter 7 of Gelman et al.\ (2003) for a
fully Bayesian treatment of finite-sample inference.} From a Bayesian decision-theoretic viewpoint, this is
even more dubious, since the loss function (\ref{loss}) is difficult to interpret and to justify.

\begin{kvothe} 
``Without a specific alternative, the best we can do is to make
posterior probability statements about $\mu$ and transfer these to the
posterior distribution of the likelihood ratio (..) 
%
There cannot be strong evidence in favor of a point null hypothesis against a general alternative
hypothesis." \SI, pages 42--44
\end{kvothe}

We further note that, once \SI~has set the principle of using the posterior distribution of the likelihood
ratio (or rather of the divergence difference since this is at least symmetric in both hypotheses), there is a
whole range of outputs available including confidence intervals on the difference, for checking whether or not
they contain zero. From our (Bayesian) perspective, this solution (a) is not Bayesian for reasons exposed
above, (b) is not parameterization invariant, and (c) relies once again on an arbitrary confidence level.

\section{Misrepresentations}\label{mismis}

We have focused in this review on Aitkin's proposals rather than on his characterizations of other statistical
methods.  In a few places, however, we believe that there have been  some
unfortunate confusions from his part.

On page 22, Aitkin describes Bayesian posterior distributions as ``formally a measure of personal uncertainty
about the model parameter,'' a statement that we believe holds generally only under a definition of ``personal''
that is so broad as to be meaningless.  As we have discussed elsewhere (Gelman, 2008), Bayesian probabilities
can be viewed as ``subjective'' or ``personal'' but this is not necessary. Or, to put it another way, if you want
to label my posterior distribution as ``personal'' because it is based on my personal choice of prior
distribution, you should also label inferences from the proportional hazards model as ``personal'' because it is
based on the user's choice of the parameterization of Cox (1972); you should also label any linear regression
(classical or otherwise) as ``personal'' as based on the individual's choice of predictors and assumptions of
additivity, linearity, variance function, and error distribution; and so on for all but the very simplest
models in existence.

In a nearly century-long tradition in statistics, any probability model is sharply divided into ``likelihood''
(which is considered to be objective and, in textbook presentations, is often simply given as part of the
mathematical specification of the problem) and ``prior'' (a dangerously subjective entity to which the
statistical researcher is encouraged to pour all of his or her pent-up skepticism).  This may be a tradition
but it has no logical basis.  If writers such as Aitkin wish to consider their likelihoods as objective and
consider their priors as subjective, that is their privilege.  But we would prefer them to restrain themselves
when characterizing the models as others.  It would be polite to either tentatively accept the objectivity of
others' models or, contrariwise, to gallantly affirm the subjectivity of one's own choices.

Aitkin also mischaracterizes hierarchical models, writing
``It is important not to interpret the prior as in some sense a {\em model for nature} [italics in the
original] that nature has used a random process to draw a parameter value from a higher distribution of
parameter values \dots''
On the contrary, that is exactly how
we interpret the prior distribution in the ideal case.  Admittedly, we do not generally approach this ideal
(except in settings such as genetics where the population distribution of parameters has a clear sampling
distribution), just as in practice the error terms in our regression models do not capture the true
distribution of errors.  Despite these imperfections, we believe that it can often be helpful to interpret the
prior as a model for the parameter-generation process and to improve this model where appropriate.

\section{Contributions of the book}\label{value}

\SI~points out several important facts that are individually known well (but perhaps not well enough!), but by
putting them all in one place it foregrounds the difficulty or impossibility of putting all the different
approaches to model checking in one place.  We all know that the $p$-value is in no way the posterior probability
of a null hypothesis being true; in addition, Bayes factors as generally practiced correspond to no actual
probability model.  Also, it is well-known that the so-called harmonic mean approach to calculating Bayes
factors is inherently unstable, to the extent that in the situations where it does ``work," it works by
implicitly integrating over a space different from that of its nominal model.

Yes, we all know these things, but as is often the case with scientific anomalies, they are associated with
such a high level of discomfort that many researchers tend to forget the problems or try to finesse them.  It
is refreshing to see the anomalies laid out so clearly.

At some points, however, Aitkin disappoints.  For example, at the end of Section 7.2, he writes:  ``In the
remaining sections of this chapter, we first consider the posterior predictive $p$-value and point out
difficulties with the posterior predictive distribution which closely parallel those of Bayes factors.''  He
follows up with a section entitled ``The posterior predictive distribution,'' which concludes with an example
that he writes ``should be a matter of {\em serious} concern [emphasis in original] to those using posterior
predictive distributions for predictive probability statements.''

What is this example of serious concern?  It is an imaginary problem in which
he observes 1 success in 10 independent trials and then is asked to compute the
probability of getting at most 2 successes in 20 more trials from the same
process.  \SI~assumes a uniform prior distribution on the success probability
and yields a predictive probability or 0.447, which, to him, ``looks a vastly
optimistic and unsound statement.'' Here, we think Aitkin should take Bayes a
bit more seriously.  If you think this predictive probability is unsound, there
should be some aspect of the prior distribution or the likelihood that is
unsound as well.  This is what Good (\citeyear{good:1950}) called ``the
device of imaginary results.''  We suggest that, rather than abandoning highly
effective methods based on predictive distributions, Aitkin should look more
carefully at his predictive distributions and either alter his model to fit his
intuitions, alter his intuitions to fit his model, or do a bit of both.  This is the value of inferential coherence as an ideal.

\section{Solving non-problems}\label{sec:noval}

Several of the examples in \SI~represent solutions to problems that seem to us to be artificial or
conventional tasks with no clear analogy to applied work.

\begin{kvothe}``They are artificial and are expressed in terms of a survey of 100 individuals expressing support
(Yes/No) for the president, before and after a presidential address (...) The question of interest is whether
there has been a change in support between the surveys  (...). We want to assess the evidence for the
hypothesis of equality $H_1$ against the alternative hypothesis $H_2$ of a change." \SI, page 147\end{kvothe}

Based on our experience in public opinion research, this is not a real question.  Support for any political
position is always changing.  The real question is how much the support has changed, or perhaps how this change
is distributed across the population.

A defender of Aitkin (and of classical hypothesis testing) might respond at this point that, yes, everybody
knows that changes are never exactly zero and that we should take a more ``grown-up'' view of the null
hypothesis, not that the change is zero but that it is nearly zero.  Unfortunately, the metaphorical
interpretation of hypothesis tests has problems similar to the theological doctrines of the Unitarian church.
Once you have abandoned literal belief in the Bible, the question soon arises:  why follow it at all?
Similarly, once one recognizes the inappropriateness of the point null hypothesis, it makes more sense not to
try to rehabilitate it or treat it as treasured metaphor but rather to attack our statistical problems
directly, in this case by performing inference on the change in opinion in the population.

To be clear:  we are not denying the value of hypothesis testing.  In this example, we find it completely
reasonable to ask whether observed changes are statistically significant, i.e. whether the data are consistent with
a null hypothesis of zero change.  What we do not find reasonable is the statement that ``the question of
interest is whether there has been a change in support.''

\begin{figure}[hbtp]
\centering
\includegraphics[width=.49\textwidth]{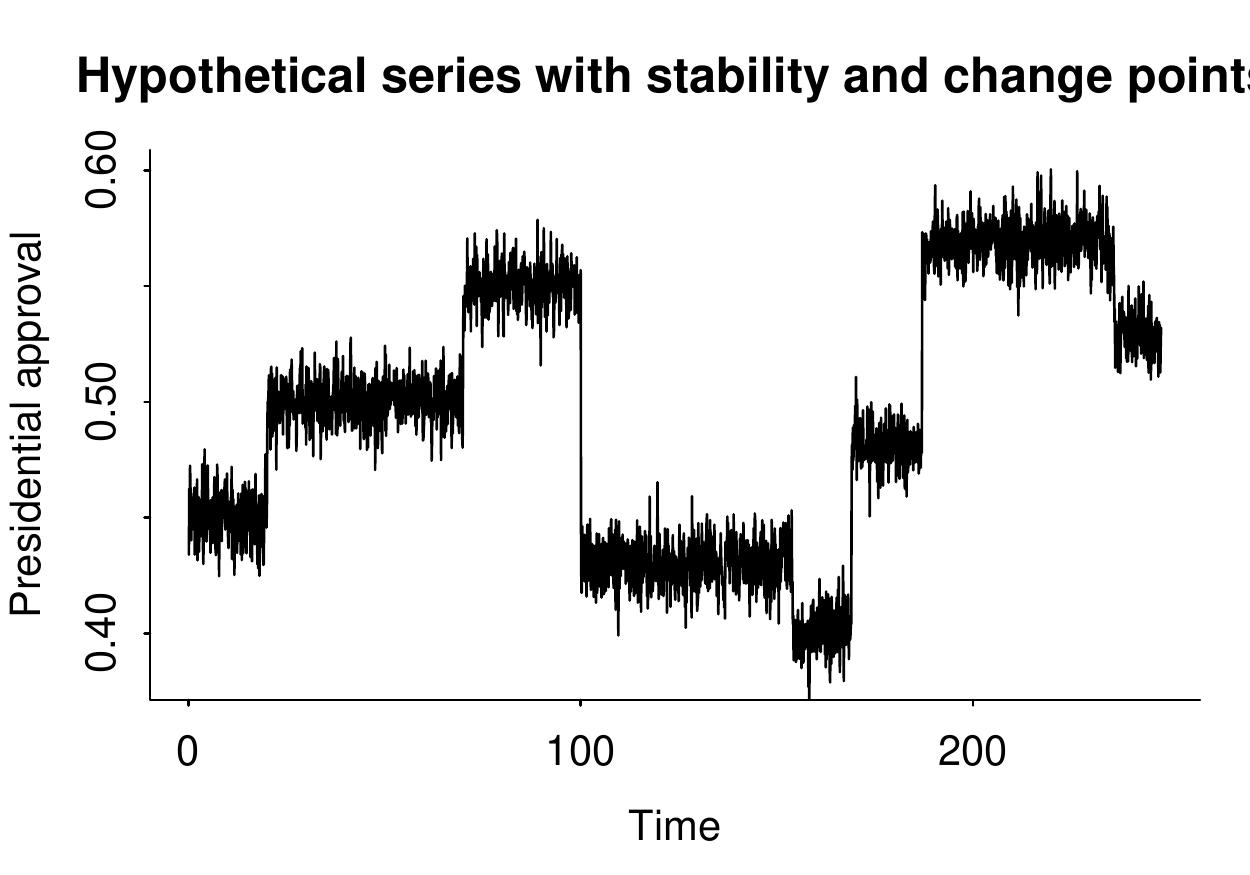}
\includegraphics[width=.49\textwidth]{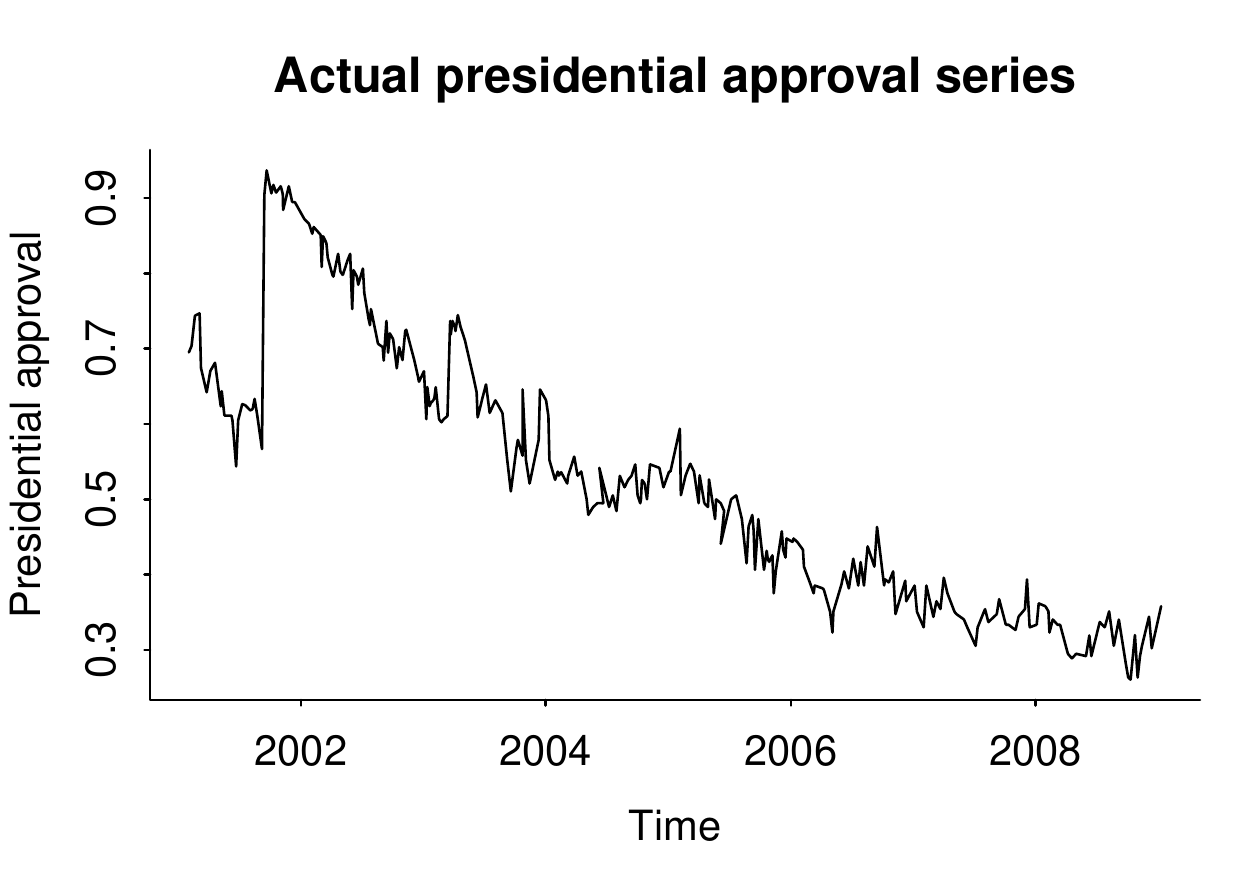}
\caption{\label{fig:president}
{\it (a) Hypothetical graph of presidential approval with discrete jumps; (b)  presidential approval series
(for George W. Bush) showing movement at many different time scales.  If the approval series looked like the
graph on the left, then Aitkin's ``question of interest" of ``whether there has been a change in support between
the surveys" would be completely reasonable.  In the context of actual public opinion data, the question does
not make sense; instead, we prefer to think of presidential approval as a continuously-varying process. }}
\end{figure}

All this is application-specific.  Suppose public opinion was observed to really be flat, punctuated by
occasional changes, as in the left graph in Figure \ref{fig:president}.  In that case, Aitkin's question of ``whether
there has been a change'' would be well-defined and appropriate, in that we could interpret the null hypothesis
of no change as some minimal level of baseline variation.

Real public opinion, however, does not look like baseline noise plus jumps, but rather shows continuous
movement on many time scales at once, as can be seen from the right graph in Figure \ref{fig:president}, which shows actual
presidential approval data.  In this example, we do not see Aitkin's question as at all reasonable.  Any attempt to work
with a null hypothesis of opinion stability will be inherently arbitrary.  It would make much more sense to
model opinion as a continuously-varying process.

The statistical problem here is not merely that the null hypothesis of zero change is nonsensical; it is that
the null is in no sense a reasonable approximation to any interesting model.  The sociological problem is that,
from \cite{savage:1954} onward, many Bayesians have felt the need to mimic the classical null-hypothesis testing
framework, even where it makes no sense.  Aitkin is unfortunately no exception, taking a straightforward
statistical question---estimating a time trend in opinion---and re-expressing it as an abstracted hypothesis
testing problem that pulls the analyst away from any interesting political questions.

\section{Conclusion:  Why did we write this review?}
\begin{kvothe}
``The posterior has a non-integrable spike at zero. This is equivalent to assigning zero prior probability to
these unobserved values.'' \SI, page 98
\end{kvothe}

%

A skeptical (or even not so skeptical) reader might at this point ask, Why did we bother to write a detailed
review of a somewhat obscure statistical method that we do not even like?  Our motivation surely was not to
protect the world from a dangerous idea; if anything, we suspect our review will interest some readers who
otherwise would not have heard about the approach (as previously illustrated by \citealp{robert:2010b}).

In 1970, a book such as \SI~could have had a large influence in statistics.  As Aitkin notes in his
preface, there was a resurgence of interest in the foundations of statistics around that time, with Lindley,
Dempster, Barnard, and others writing about the intersections between classical and Bayesian inference (going
beyond the long-understood results of asymptotic equivalence) and researchers such as Akaike and Mallows
beginning to integrate model-based and predictive approaches to inference.  A glance at the influential text of
Cox and Hinkley (1974) reveals that theoretical statistics at that time was focused on inference from
independent data from specified sampling distributions (possibly after discarding information, as in rank-based
tests), and ``likelihood'' was central to all these discussions.

Forty years on, a book on likelihood inference is more of a niche item.  Partly this is simply part of the
growth of the field---with the proliferation of books, journals, and online publications, it is much more
difficult for any single book to gain prominence.  More than that, though, we think statistical theory has moved
away from iid analysis, toward more complex, structured problems.

%

That said, the foundational problems that \SI~discusses are indeed important and they have not yet been
resolved.  As models get larger, the problem of ``nuisance parameters'' is revealed to be not a mere nuisance
but rather a central fact in all methods of statistical inference.  As noted above, Aitkin makes valuable
points---known, but not well-enough known---about the difficulties of Bayes factors, pure likelihood, and other
superficially attractive approaches to model comparison.  We believe it is a natural continuation of this work
to point out the problems of the integrated likelihood approach as well.

For now, we recommend model expansion, Bayes factors where reasonable, cross-validation, and predictive model
checking based on graphics rather than $p$-values.  We recognize that each of these approaches has loose ends.
But, as practical idealists, we consider inferential challenges to be opportunities for model improvement with
the Bayesian realm rather than motivations for a new theory of noninformative priors that takes us in uncharted territories.


\end{document}